\documentclass[aps,pra,reprint, amsmath, amssymb,superscriptaddress,nofootinbib]{revtex4-1}

\usepackage{bm}
\usepackage[retainorgcmds]{IEEEtrantools}
\usepackage{graphicx}
\usepackage{mathrsfs}
\usepackage{amsmath}
\usepackage{amssymb}
\usepackage{color}
\usepackage{amsfonts}
\usepackage{times,txfonts}
\usepackage{nicefrac}
\usepackage[colorlinks=true,linkcolor=blue,urlcolor=blue,citecolor=blue,pdfusetitle]{hyperref}
\usepackage{amsmath}
\DeclareMathOperator{\sign}{sign}

\newcommand{\tr}{\text{tr}}

\begin{document}

\title{Limiting flux in quantum thermodynamics}
\date{\today}
\author{Domingos S. P. Salazar}
\affiliation{Unidade de Educa\c c\~ao a Dist\^ancia e Tecnologia,
Universidade Federal Rural de Pernambuco,
52171-900 Recife, Pernambuco, Brazil}

\begin{abstract}
In quantum systems, entropy production is typically defined as the quantum relative entropy between two states. This definition provides an upper bound for any flux (of particles, energy, entropy, etc.) of bounded observables, which proves especially useful near equilibrium. However, this bound tends to be irrelevant in general nonequilibrium situations. We propose a new upper bound for such fluxes in terms of quantum relative entropy, applicable even far from equilibrium and in the strong coupling regime. Additionally, we compare this bound with Monte Carlo simulations of random qubits with coherence, as well as with a model of two interacting nuclear spins.
\end{abstract}
\maketitle{}

{\bf \emph{Introduction -}} The concept of entropy production is pivotal in understanding nonequilibrium phenomena \cite{RevModPhys.93.035008,Campisi2011,Seifert2012Review,Batalhao2015,Campisi2011,Esposito2009,Ciliberto2013,Batalhao2014,Barato2015A,Gingrich2016,Pietzonka2017,VanTuan2020,Liu2020,Horowitz2022,Koyuk2019,Dechant2023,Mori2023,Zhang2023}. It is typically associated with various types of fluxes (energy, particles, entropy, etc.), especially near equilibrium where Onsager's relations are significant \cite{Onsager1931}. In quantum systems, flux is often expressed using an observable $\hat{\theta}$ as $\phi:=\tr (\hat{\theta}(\rho-\sigma))$ between states $\sigma$ and $\rho$. For bounded operators, a frequent scenario in finite size reservoirs, the absolute flux is limited by a capacity $\phi_L$, 
\begin{equation}
\label{flux}
|\phi| \leq \phi_L:=\theta_{max} -\theta_{min}, 
\end{equation}
where $\theta_{max}:=\sup_n \theta_n$, $\theta_{min} := \inf_n \theta_n$ and $\{\theta_n\}$ are the eigenvalues of the bounded operator $\hat{\theta}$, and it follows that $(\phi/\phi_L)^2 \leq 1$. Intuitively, the maximum flux $\phi_L$ corresponds to an exchange between two pure states $|\theta_{min}\rangle$ and $|\theta_{max}\rangle$.
However, in practical situations, this capacity is seldom reached. In fact, if $\rho$ and $\sigma$ are known, a more effective upper bound for $|\phi|/\phi_L$ can be derived in terms of the quantum relative entropy $S(\rho||\sigma):=\tr(\rho(\ln \rho - \ln \sigma))$, linked to quantum entropy production \cite{RevModPhys.93.035008}. Controlling such absolute flux (\ref{flux}) through entropy production in quantum systems could serve as a fundamental tool, similar to the Thermodynamic Uncertainty Relations (TURs) \cite{Barato2015A,Gingrich2016,Polettini2017,Pietzonka2017,Hasegawa2019,Hasegawa2019b,VanTuan2020,Van_Vu_2020,
Timpanaro2019b,Liu2020,Horowitz2022,Potts2019,Proesmans2019,Salazar2022d,Diebal2023,TanVanVu2023,Yoshimura2023}.

Interestingly, thermodynamics limits the flux by means of the entropy production in the same way that quantum mutual information is used as a bound for connected correlators \cite{Wolf2008,Kudler2023}.
To see this, first we define the Schatten $k-$norm $||x||_k:=\tr[(\sqrt{x^\dagger x})^k]^{1/k}$, which results in the trace norm for the particular case $k=1$. Then, we have $|\phi| = |\tr((\hat{\theta}-\lambda I)(\rho-\sigma))| \leq || \hat{\theta} (\rho-\sigma) ||_1$, where $\lambda:=(\theta_{max}+\theta_{min})/2$ and $I$ is the identity operator. Now using H\"{o}lder's inequality, $||(\rho-\sigma)(\hat{\theta}-\lambda I)||_1 \leq ||\rho-\sigma||_1 ||\hat{\theta}-\lambda I||_\infty$, it finally results in
\begin{equation}
\label{firstbound}
   \Big(\frac{ \phi}{\phi_L}\Big)^2 \leq \frac{||\rho - \sigma||_1^2}{4}\leq \frac{S(\rho||\sigma)}{2},
\end{equation}
after using $||\hat{\theta} - \lambda I||_\infty^2 = \sup_n |\theta_n - \lambda|^2= \phi_L^2/4$ and the quantum Pinsker's inequality, $S(\rho||\sigma)\geq (1/2)||\rho-\sigma||_1^2$.

The connection of relation (\ref{firstbound}) with thermodynamics goes as follows. When system (S) and the environment (E) are initially independent and prepared in arbitrary states, $\rho_{SE}=\rho_S \otimes \rho_E$, then a unitary dynamics $\mathcal{U}$ potentially builds classic and quantum correlation between them, resulting in the final state $\rho_{SE}'=\mathcal{U}(\rho_S \otimes \rho_E)\mathcal{U}^\dagger$. In this notation, the quantum entropy production \cite{Esposito2010a,Manzano2018PRX,Jarzynski2011a,Deffner2011,Goold2016,Elouard2023} is defined as $\Sigma :=S(\rho_{SE}'||\rho_S' \otimes \rho_E)$,
where $\rho_S':=\tr_E(\rho_{SE}')$ is the reduced state of the system. In this case, we have from (\ref{firstbound}) with $\rho=\rho_{SE}'$ and $\sigma=\rho_S' \otimes \rho_E$,
\begin{equation}
\label{thermobound}
   \Big(\frac{ \phi}{\phi_L}\Big)^2 \leq \frac{\Sigma}{2}.
\end{equation}
Expression (\ref{thermobound}) proves particularly useful for small values of $\Sigma$, resembling Onsager's relation near equilibrium. Onsager's relation, based on linear response theory, is characterized by a quadratic form connecting fluxes and entropy production \cite{DeGroot1961}. However, this quadratic expression $\Sigma \propto \phi^2$ becomes invalid far from equilibrium, where the relationship between entropy production and fluxes is only applicable to specific systems \cite{SalazarLandi2020}. Notably, the bound (\ref{thermobound}) also loses its relevance for $\Sigma \geq 2$, a scenario indicative of being far from equilibrium. In such a region, Pinsker's inequality becomes inapplicable, and a more fitting bound for (\ref{thermobound}) would be the straightforward $(\phi/\phi_L)^2\leq 1 \leq \Sigma/2$, as derived from (\ref{flux}).

In this letter, we explore whether entropy production can still be a tool for controlling fluxes in conditions far from equilibrium. We affirmatively answer this with our main result,
\begin{equation}
\label{main}
  \Big(\frac{\phi}{\phi_L}\Big)^2 \leq B\Big(\frac{S(\rho||\sigma)+S(\sigma||\rho)}{2}\Big)\leq 1,
\end{equation}
where $B(x):=(x/g(x))^2$ and $g(x)$ is the inverse of $h(x)=x\tanh(x/2)$ for $x>0$. Expression (\ref{main}) means that, even arbitrarily far from equilibrium, the quantum relative entropy can be used as bound for limiting fluxes. In quantum thermodynamics, setting again $\rho=\rho_{SE}'$ and $\sigma=\rho_S'\otimes \rho_E$, we get $\Sigma=S(\rho||\sigma)$ and we define the dual $\Sigma^*:=S(\sigma||\rho)=S(\mathcal{U^\dagger}\sigma\mathcal{U}||\rho_S \otimes \rho_E)$. Note that $\Sigma^*$ is uniquely defined from the same states used in $\Sigma$ \cite{Salazar2023d}. Moreover, relation (\ref{main}) is specially useful for systems arbitrarily far from equilibrium, where (\ref{thermobound}) fails to bring any information (for $\Sigma \geq 2$) and (\ref{main}) always produces a nontrivial bound. In this case, we also obtain a relationship between the entropy production and any flux from (\ref{main}),
\begin{equation}
\label{Onsagerslike2}
\frac{\Sigma+\Sigma^*}{4} \geq \frac{\phi}{\phi_L} \tanh^{-1} \big( \frac{\phi}{\phi_L}\big) \geq \big(\frac{\phi}{\phi_L}\big)^2,
\end{equation}
which is our second main result, valid for bounded operators $\hat{\theta}$, including for the entropy flux (when $\hat{\theta}=\log \rho_E$). 

The letter is organized as follows: First, we establish the formalism and prove the main results (\ref{main}) and (\ref{Onsagerslike2}). We also test (\ref{main}) in a simulation involving two interacting qubits, with one acting as the system and the other as the environment. As an example, we demonstrate how the energy flux is controlled by the quantum relative entropy in a model of two interacting nuclear spins-1/2.

{\bf \emph{Formalism -}} We prove our main result (\ref{main}) using a recently proposed quantum thermodynamic uncertainty relation (qTUR) \cite{Salazar2023d}. For any Hermitian operator $\hat{\omega}$ and states $\rho$, $\sigma$ (with $\langle \hat{\omega} \rangle_\rho \neq \langle \hat{\omega} \rangle_\sigma$), the qTUR states that
\begin{equation}
\label{form1}
\frac{\langle \hat{\omega}^2 \rangle_\rho - \langle \hat{\omega} \rangle_\rho^2 + \langle \hat{\omega}^2 \rangle_\sigma - \langle \hat{\omega} \rangle_\sigma^2}{(1/2)(\langle \hat{\omega} \rangle_\rho - \langle \hat{\omega} \rangle_\sigma)^2} \geq f(\frac{S(\rho||\sigma) + S(\sigma||\rho)}{2}),
\end{equation}
where $f(x)=1/\sinh(g(x)/2)^2$ and the notation $\langle \hat{\omega} \rangle_{\rho}:=\tr(\hat{\omega}\rho)$. Now we consider a specific operator $\hat{\omega}$ and calculate the averages $\langle \hat{\omega} \rangle_{\rho,\sigma}$, $\langle \hat{\omega}^2 \rangle_{\rho,\sigma}$ in the lhs of (\ref{form1}). The idea is to select $\hat{\omega}$ such that the trace norm $||\rho-\sigma||_1$ appears naturally. 

First, note that $\rho-\sigma$ is Hermitian and it has a decomposition with real eigenvalues, $\rho-\sigma = \sum_k w_k |w_k \rangle \langle w_k |$. Now we define $\hat{\omega}$ as
\begin{equation}
\label{form2}
\hat{\omega}:= \sum_{k, w_k \neq 0} \sign(w_k) |w_k\rangle \langle w_k|,
\end{equation}
where $\sign(x)=1~(-1)$, for $x>0~(x<0)$. Then, we write the trace-norm $||\rho-\sigma||_1$ in terms of $\hat{\omega}$ from (\ref{form2}),
\begin{equation}
\label{form3}
\langle \hat{\omega} \rangle_\rho - \langle \hat{\omega} \rangle_\sigma = \tr[\hat{\omega}(\rho-\sigma)]=\sum_k |w_k| =||\rho - \sigma||_1,
\end{equation}
which meets the condition for the qTUR (\ref{form1}), $\langle \hat{\omega}\rangle_\sigma \neq \langle \hat{\omega}\rangle_\sigma$, for $\rho \neq \sigma$. Then, we observe that 
\begin{equation}
\label{form4}
\hat{\omega}^2 = \sum_{k,w_k \neq 0} \sign(w_k)^2 |w_k\rangle \langle w_k| = I - \hat{\epsilon},
\end{equation}
where $I$ is the identity operator and $\hat{\epsilon}:=\sum_{k,w_k=0}|w_k\rangle \langle w_k|$ , with averages
\begin{equation}
\label{form5}
\langle \hat{\epsilon} \rangle_\rho = \langle \hat{\epsilon} \rangle_\sigma := \epsilon,
\end{equation}
obtained from $\langle \hat{\epsilon}\rangle_\rho - \langle \hat{\epsilon}\rangle_\sigma = \tr[\hat{\epsilon}(\rho-\sigma)] = \sum_{k,w_k=0} w_k = 0$. We also have $0 \leq \epsilon \leq 1$, because $\rho, \sigma$ are positive definite and $\tr(\rho)=\tr(\sigma)=1$. From (\ref{form4}) and (\ref{form5}), we get
\begin{equation}
\label{form7}
\langle \hat{\omega}^2 \rangle_\rho = \langle \hat{\omega}^2\rangle_\sigma = 1 - \epsilon.
\end{equation}
Using the averages (\ref{form3}) and (\ref{form7}) in 
the qTUR (\ref{form1}), we obtain
\begin{equation}
\label{form8}
\frac{2-2\epsilon - \langle \hat{\omega}\rangle_\rho^2 - \langle \hat{\omega}\rangle_\sigma^2}{(1/2)||\rho-\sigma||_1^2} \geq f(\tilde{S}(\rho,\sigma)),
\end{equation}
with the notation $\tilde{S}(\rho||\sigma):=(S(\rho||\sigma)+S(\sigma||\rho))/2$. As third and final ingredient, check that
\begin{equation}
\label{form9}
 (1/2)||\rho - \sigma ||_1^2 \leq \langle \hat{\omega} \rangle_\rho^2 + \langle \hat{\omega}\rangle_\sigma^2,
\end{equation}
directly from (\ref{form3}) and the expression $(1/2)(x-y)^2 \leq  (1/2)(x-y)^2 + (1/2)(x+y)^2 = x^2 + y^2$, for $x=\langle \hat{\omega} \rangle_\rho$, $y=\langle \hat{\omega} \rangle_\sigma$. In this case, we obtain from (\ref{form9}),
\begin{equation}
\label{form10}
  \frac{2-2\epsilon - (1/2)||\rho-\sigma||_1^2}{(1/2)||\rho-\sigma||_1^2} \geq  \frac{2-2\epsilon - \langle \hat{\omega}\rangle_\rho^2 -\langle \hat{\omega}\rangle_\sigma^2}{(1/2)||\rho-\sigma||_1^2}.
\end{equation}
Combining (\ref{form10}) and (\ref{form8}) results in
\begin{equation}
\label{form11}
 \frac{2-2\epsilon - (1/2)||\rho-\sigma||_1^2}{(1/2)||\rho-\sigma||_1^2} \geq f(\tilde{S}(\rho,\sigma)).
\end{equation}
Expression (\ref{form11}) is easily inverted to
\begin{equation}
\label{form12}
    \frac{||\rho - \sigma||_1^2}{4} \leq \frac{(1-\epsilon)}{1+f(\tilde{S}(\rho,\sigma))} = (1-\epsilon)
    B\big(\tilde{S}(\rho,\sigma)\big),
\end{equation}
with $B(x):=(x/g(x))^2$. Finally, using the first inequality of (\ref{firstbound}), we obtain $(\phi/\phi_L)^2 \leq (1-\epsilon)B(\tilde{S}(\rho,\sigma)) \leq B(\tilde{S}(\rho,\sigma))$, which is our main result (\ref{main}). Check that $B(x)=(x/g(x))^2 \leq 1$, for $x>0$, because $x\tanh(x/2)=h(x) \leq x = h(g(x))$, which makes $x \leq g(x)$, because $h$ is a increasing function.

{\bf \emph{Discussion-}} We proved a stronger result in terms of $\epsilon$ in (\ref{form12}) that reproduces the main result (\ref{main}) in case $\epsilon=0$. Although we used the qTUR (\ref{form1}) in the proof, a similar reasoning was used in recent classic results \cite{Salazar2022b,Vo_2022,Dechant2022}, connecting the total variation distance (TV) and the symmetric Kullback-Leibler (KL) divergence. Actually, if $[\rho,\sigma]=0$, relation (\ref{form12}) reduces to a upper bound for the TV in terms of the symmetric KL divergence. In this sense, our quantum result (\ref{main}) generalizes previous classic results in information theory and stochastic thermodynamics.

As in the classic case, the bound in (\ref{form12}) is saturated for a specific minimal two-level system. Consider $\rho=[e^{a/2}|1\rangle \langle 1| + e^{-a/2}|0\rangle \langle 0|]/(2\cosh(a/2))$,  $\sigma=[e^{-a/2}|1\rangle \langle 1| + e^{a/2}|0\rangle \langle 0|]/(2\cosh(a/2))$. In this case, one has $||\rho-\sigma||_1/2 = \tilde{S} :=[S(\rho||\sigma)+S(\sigma||\rho)]/2=\tanh(|a|/2)$ and $\tilde{S}^2/g(\tilde{S})^2=\tanh(a/2)^2$ and $\epsilon=0$. Therefore, $||\rho-\sigma||_1^2/4 = \tilde{S}^2/g(\tilde{S})^2=B(\tilde{S})$ saturates (\ref{form12}).

Before moving to the applications, it is worthy to mention that the choice $\lambda=|\theta_{max}+\theta_{min}|/2$ in the derivation of (\ref{firstbound}) was selected for a reason.
If we introduce a shift in the operator given by $\hat{\theta}\rightarrow \hat{\theta}-\lambda I$, where $I$ is the identity and $\lambda\in \mathbb{R}$, we note that $\phi=\tr((\hat{\theta}-\lambda I)(\rho-\sigma))=\tr(\hat{\theta}(\rho-\sigma))$. It means that the flux is invariant under the shift. In this case, one could rewrite (\ref{main})
for any $\lambda$ and we could search the optimal value $\lambda^*$ for a given observable $\hat{\theta}$ so that the lhs of (\ref{firstbound}) is tighter. In other words, we want to minimize $||\hat{\theta}-\lambda I ||_\infty$. Formally, we obtain
\begin{equation}
\label{disc2}
\inf_{\lambda \in \mathbb{R}} ||\hat{\theta}-\lambda I||_\infty = \frac{1}{2}(\theta_{max} - \theta_{min}) = \frac{\phi_L}{2}.
\end{equation}
The proof of (\ref{disc2}) follows from $||\hat{\theta} - \lambda I||_\infty = \sup_n |\theta_n - \lambda| = \max(|\theta_{max}-\lambda|;|\theta_{min}-\lambda|)$, then using $\inf_\lambda \max(|\theta_{max}-\lambda|;|\theta_{min}-\lambda|) = (\theta_{max}-\theta_{min})/2$ and the optimal value $\lambda^*=(\theta_{max}+\theta_{min})/2$. In this case, we conclude that the bound (\ref{firstbound}) is actually the tightest for any shift,
\begin{equation}
\label{main2}
  \frac{\phi^2}{4||\hat{\theta}-\lambda I||_\infty^2} \leq\big(\frac{\phi}{\phi_L}\big)^2 \leq \frac{||\rho-\sigma||_1^2}{4},
\end{equation}
for any $\lambda \in \mathbb{R}$, which makes (\ref{thermobound}) and (\ref{main}) particularly useful because they are now invariant under shifts of the operator $\hat{\theta}$.

{\bf \emph{Simulations -}} We now test numerically our main result (\ref{main}) for a very simple system: two random qubits $\rho$, $\sigma$ including quantum coherence. For each run, we draw random states $\rho,\sigma$ and a random operator $\hat{\theta}$. We compute the flux $\phi=\tr(\hat{\theta}(\rho-\sigma))$ and $\phi_L$ using (\ref{flux}). 
\begin{figure}[htp]
\includegraphics[width=3.3 in]{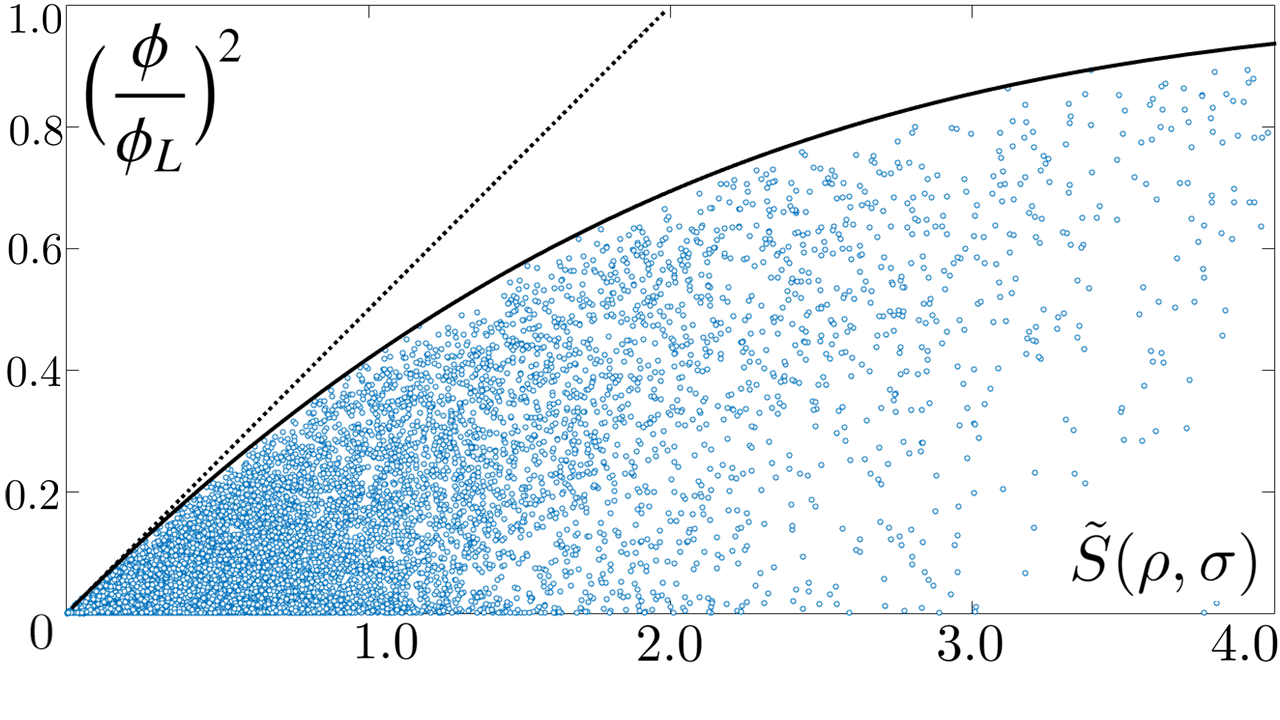}
\caption{(Color online) Monte Carlo simulation of the the flux $\phi$ for random observables $\hat{\theta}$ and states $\rho$, $\sigma$ as a function of the symmetric quantum relative entropy $\tilde{S}(\rho,\sigma)=[S(\rho||\sigma)+S(\sigma||\rho)]/2$, for $n=10^4$ draws of ($\hat{\theta},\rho,\sigma$). The dashed line represents the bound $(\phi/\phi_L)^2 \leq \tilde{S}(\rho,\sigma)/2$ which works particularly well near equilibrium, but it starts to depart significantly from the blue points for $\tilde{S}\gg 0$. The solid line represents the bound $(\phi/\phi_L)^2 \leq B(\tilde{S}(\rho,\sigma))$, which is our main result. Note that the solid line manages to produce a nontrivial bound arbitrarily far from equilibrium and, for $\tilde{S}(\rho,\sigma)=2$, the dashed line fails to produce a meaningful bound, as it hits the maximum $(\phi/\phi_L)^2=1.$
}
\label{fig1}
\end{figure}
We also compute the quantum relative entropies $S(\rho||\sigma)$ and $S(\sigma||\rho)$. Then, we plot $(\phi/\phi_L)^2$ vs. $\tilde{S}(\rho,\sigma):=[S(\rho||\sigma)+S(\sigma+\rho)]/2$. In Fig.~1, we compare two bounds: the second inequality of (\ref{firstbound}) in the dashed line (quadratic relationship, similar to Onsager's) and our main result (\ref{main}) in the solid line. The draws were realized as follows \cite{Salazar2023d} . Let $X \sim I_x$ 
be a random variable uniformly distributed in the interval $I_x$. We consider the decomposition $\rho = (1-p_1)
|0\rangle \langle 0| + p_1 |1\rangle \langle 1|$, where $p_1 \sim [0,1]$ for each run. We independently draw a random $\sigma = (1-q_1)
|0\rangle \langle 0| + q_1 |1\rangle \langle 1| + C|0\rangle \langle 1| + C^*|1\rangle \langle 0|$, where $q_1 \sim [0,1]$, with $C:=|C|\exp(\phi_1 i)$, where $|C|^2 \sim [0,q_1(1-q_1)]$, $\phi_1 \sim [0,2\pi)$, so that $\sigma$ is completely positive. Finally, we draw a random Hermitian operator $\hat{\theta}=\omega(|1\rangle \langle 1| - |0\rangle \langle 0|) + D|0\rangle \langle 1| + D^*|1\rangle \langle 0|$, where $\hat{\omega} \sim [0,4]$ and $D:=|D|\exp(\phi_2 i)$, with $|D|^2 \sim [0,1]$, $\phi_2 \sim [0,2\pi)$. The pairs [$(\phi/\phi_L)^2, \tilde{S}(\rho,\sigma$)] are depicted as blue points for $n=10^4$ runs. Note that the dashed curve becomes innocuous for $\tilde{S}(\rho,\sigma)=2$, where the solid line produces a useful bound arbitrarily far from equilibrium.

{\bf \emph{Application: local operators and entropy flux -}} 
Now we apply (\ref{main}) for a specific choice of local operator $\hat{\theta}$ acting on the environment. The choice $\hat{\theta}=\log\rho_E$ in $(\ref{main})$ results in $\phi=\Phi:=\tr_E((\rho_E -\rho_E')\log \rho_E)$, which is the quantum entropy flux \cite{RevModPhys.93.035008}, with $\rho=\rho_{E}'$ and $\sigma=\rho_E$. A direct application of (\ref{main}) yields
\begin{equation}
\label{Onsagerslike1}
\big(\frac{\Phi}{\Phi_L}\big)^2 \leq B\big(\tilde{S}(\rho_E',\rho_E)\big) \leq 1,
\end{equation}
where (\ref{Onsagerslike1}) is valid
for bounded environments $||\ln \rho_E||_\infty < \infty$. From (\ref{Onsagerslike1}), after using $B(x)=(x/g(x))^2=[\tanh(g(x)/2)]^2$ and the data processing inequality $S(\rho||\sigma)\geq S(\varepsilon(\rho)||\varepsilon(\sigma))$, where $\varepsilon$ is a CPTP map, we also obtain,
\begin{figure}[htp]
\includegraphics[width=3.3 in]{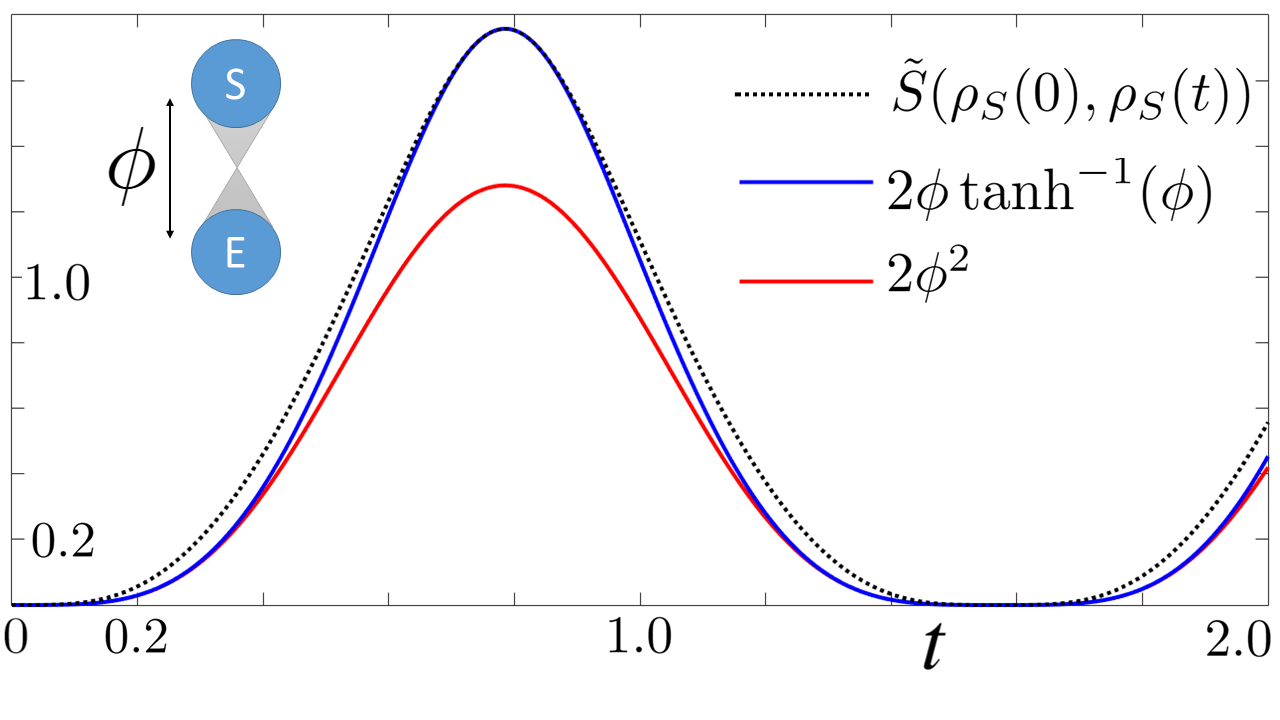}
\caption{(Color online) Simulation of two interacting qubits initially uncorrelated. The dynamics builds correlation over time, but it preserves the total energy of the system. An energy flux $\phi$ (where $\phi_L=1$) is observed between $S$ and $E$, whose absolute value is displayed in two different forms: quadratic ($2\phi^2$, in red) and our result ($2\phi\tanh^{-1}(\phi)$, in blue). We also show the symmetric quantum relative entropy $\tilde{S}(\rho_S(0),\rho_S(t))$ in terms of the reduced states as a function of time (dotted line). We see that the quadratic form (red) and our result (blue) are initially close, because the flux is small. Over time, as the flux is not negligible, the quantum relative entropy approaches the blue line.
}
\label{fig2}
\end{figure}
\begin{equation}
\label{EntropyFlux1}
\frac{\Sigma+\Sigma^*}{2} \geq \tilde{S}(\rho_E,\rho_E')\geq 2\frac{\Phi}{\Phi_L} \tanh^{-1} \big( \frac{\Phi}{\Phi_L}\big) \geq 2\big(\frac{\Phi}{\Phi_L}\big)^2,
\end{equation}
a general relation between entropy production and flux that resembles Onsager's relations for small fluxes ($\Phi/\Phi_L \approx 0$, last inequality), 
but it contains higher order terms in $\Phi$ for $|\Phi|\rightarrow \Phi_L$. It is valid arbitrarily far from equilibrium in the strong coupling regime and for nonthermal environments (as long as they are bounded, $||\ln \rho_E ||_\infty < \infty$). The same reasoning is valid for any flux (not only entropy flux), thus we also have (\ref{main2}) in general. In the case of thermal environments, $\rho_E = \exp(-\beta H_E)/Z(\beta)$, with $Z(\beta):=\tr(\exp(-\beta H_E))$, then $\Phi=\beta \tr((\rho_E'-\rho_E)H_E)$ and $\Phi_L=\beta|E_{max}-E_{min}|$ in (\ref{Onsagerslike2}), where $\beta=1/(k_B T)$, $T$ is the temperature, $\{E_n\}$ are the eigenvalues of $H_E$.

Similarly, in terms of a bounded operator $\hat{\theta}_S$ that acts locally on the system, we have $\phi=\tr[\hat{\theta}_S(\rho_S(t)-\rho_S(0))]$ and one could write (\ref{EntropyFlux1}) in terms of the reduced state of the system
\begin{equation}
\label{EntropyFlux2}
\tilde{S}(\rho_S(t),\rho_S(0)) \geq 2\frac{\phi}{\phi_L} \tanh^{-1} \big( \frac{\phi}{\phi_L}\big) \geq 2\big(\frac{\phi}{\phi_L}\big)^2,
\end{equation}
where $\rho_S(t)=\tr_E(\rho(t))$, for $t>0$. Now let us verify (\ref{EntropyFlux2}) in a physical system. The goal is to show that, even when the last inequality in (\ref{EntropyFlux2}) is still useful ($\tilde{S}(\rho,\sigma)\leq 2$), the first one might be significantly better.

{\bf \emph{Example - }} Consider two qubits interacting as proposed in \cite{Micadei2017} in a experimental realization of two interacting nuclear spins-1/2. This system is interesting because, depending on the initial correlations, it might exhibit a reverse heat flow. In our case, we consider the system initially uncorrelated. The local Hamiltonians are $H_i = \Omega |e\rangle \langle e|_i$, $i=S,E$, and the system is initially prepared in a product state of the form $\rho_{SE}=\rho_S(0) \otimes \rho_E(0)$, where $\rho_S(0)= (1-p)|g\rangle\langle g| + p |e\rangle \langle e|$ and $\rho_E(0)=(1-q)|g\rangle \langle g| + q|e\rangle \langle e|$. At $t=0$, the system interacts with an energy preserving unitary dynamics $\mathcal{U}=\exp\{-igt(e^{i\omega_0}|g,e\rangle \langle e,g|+e^{-i\omega_0}|e,g\rangle\langle g,e|)\}$, where $g$ is the strength of the interaction and $\omega_0$ is an arbitrary phase. We consider the following absolute flux $|\phi(t)|=|\tr\{H_S [\rho_S(t) - \rho_S(0)]\}|=\sin(gt)^2|p-q|$ \cite{RevModPhys.93.035008}, where $\rho_S(t)=\tr_E\{\rho(t)\}$. In Fig.~2, we observe $\tilde{S}(\rho_S(t),\rho_S(0))$ in the dotted line as a function of time for $p=0.9$, $q=0.1$, $\Omega=1$ (which makes $\phi_L=1$), $g=2$, $\omega_0=0$. In this case, note that (\ref{EntropyFlux2}) is satisfied for all $t\geq0$. Particularly, the function $2\phi \tanh^{-1}(\phi)$ approaches the quantum relative entropy $\tilde{S}(\rho_S(0),\rho_S(t))$ over time, saturating the bound at $gt = \pi/2~(t\approx 0.79)$, where the quadratic function $2\phi^2$ misses the quantum relative entropy by some margin when the flux increases. This is not always the case, as depending on the choices of the parameters, we could have a very small flux $\phi \approx 0$, which makes both functions indistinguishable, $2\phi^2 \approx 2\phi\tanh^{-1}(\phi)$.

{\bf \emph{Application: Correlation function -}} Consider again the framework of quantum thermodynamics, now with a general bounded observable of the form $\hat{\theta}=\hat{\theta}_S \otimes \hat{\theta}_E$, such that $\hat{\theta}_S$ and $\hat{\theta}_E$ act on the system and the environment, respectively. We are interested in the following correlation function, 
\begin{equation}
\label{correlation}
\tr(\hat{\theta}(\rho-\sigma))=C(\hat{\theta}_S,\hat{\theta}_E):=\langle \hat{\theta}_S \otimes \hat{\theta}_E\rangle_{\rho_{SE}'}-\langle \hat{\theta}_S \rangle_{\rho_S'} \langle \hat{\theta}_E \rangle_{\rho_E},
\end{equation}
for $\rho=\rho_{SE}'$ and $\sigma=\rho_S'\otimes\rho_E$. A proper correlation function \cite{Wolf2008,Kudler2023}   would consider $\rho_E'$ instead of $\rho_E$, but we keep this form of the``bath reset'' protocol to match the definition of the entropy production $\Sigma$. As intuition, check that if $\mathcal{U}=I$, then we are left with the initial conditions, $\rho_{SE}'=\rho_S \otimes \rho_E$ and $\rho_S'=\rho_S$, which makes $C(\hat{\theta}_S,\hat{\theta}_E)=0$. But for a general dynamics $\mathcal{U}$, correlation will build up over time. Correlations are important as they are typically related to thermodynamic properties in classic systems \cite{Ohga2023}. For this choice of operator, our main result (\ref{main}) reads
\begin{equation}
\label{correlation2}
\frac{C(\hat{\theta}_S,\hat{\theta}_E)^2}{(\theta_{max}-\theta_{min})^2} \leq B(\frac{\Sigma+\Sigma^*}{2}) \leq 1.
\end{equation}
Expression (\ref{correlation}) is also easily adjustable to account for other definitions of entropy production (for instance, the ``both reset'' protocol would replace $\rho_S'$ for $\rho_S$ in (\ref{correlation})). Particularly, in the ``both reset'' protocol, $\rho=\rho_{SE}'$ and $\sigma=\rho_S \otimes \rho_E$, it makes $\Sigma^*=S(\sigma||\rho)$ physically meaningful as it is the entropy production of the backwards process.

{\bf \emph{Conclusions - }}
We have studied the flux $\phi$ of bounded operators in quantum thermodynamics. We believe that limiting the flux in terms of information theoretic quantities is in the same spirit of the TURs. To address this task, we defined the maximum flux ($\phi_L$) and analyzed the ratio $|\phi|/\phi_L$. As it turns out, this ratio is limited by the quantum relative entropy (\ref{thermobound}) (which is associated with the entropy production in quantum thermodynamics), as a consequence of the quantum Pinsker's inequality, but we argued that this bound is not very useful far from equilibrium. 

In this context, we showed an upper bound (\ref{main}) for $(\phi/\phi_L)^2$ in terms o the quantum relative entropy that is relevant in nonequilibrium situations. The result was obtained as a consequence of a recently proposed quantum TUR and it generalizes classic results from stochastic thermodynamics. We verified the bound in Monte Carlo simulations of two random qubits and random bounded operators. We also applied the main result for local operators, obtaining a relation between entropy production and a local flux (for instance, the entropy or energy fluxes). As an example, we tested the result in a model of two interacting nuclear spins-1/2 system, where the the local energy flux of the system is limited by a function of the (symmetric) quantum relative entropy as expected from our main result.

Flux limitations by information theoretic quantities provide new insights into the nature of quantum thermodynamic processes. We believe the results of this paper might impact design and understanding of quantum systems, particularly in contexts like quantum computing, quantum heat engines, or other areas where energy transfer is plays a major role.

\bibliography{Lib8}
\end{document}